\documentclass[runningheads]{llncs}

\usepackage[T1]{fontenc}
\usepackage{graphicx}
\usepackage{booktabs}  
\usepackage{tabularx}  
\usepackage{multirow}  
\usepackage{makecell}  
\usepackage{xcolor}    
\usepackage{hyperref}  
\usepackage{float}
\usepackage{siunitx}

\usepackage{pdflscape}  
\usepackage{array}   

\usepackage{subcaption} 
\usepackage[shortcuts]{extdash}

\hypersetup{
  colorlinks=true,
  linkcolor=blue,
  filecolor=magenta,
  urlcolor=blue,
  citecolor=blue,
}

\usepackage{textgreek} 
\usepackage{amssymb}  

\DeclareUnicodeCharacter{03A8}{\textPsi}  
\DeclareUnicodeCharacter{2248}{\approx}   
\DeclareUnicodeCharacter{2212}{\textrm{-}} 

\begin{document}

\title{AI Systems in Text-Based Online Counselling: Ethical Considerations Across Three Implementation Approaches}

\titlerunning{AI Systems in Text-Based Online Counselling}

\author{%
Philipp Steigerwald\inst{1}\orcidID{0009-0002-5564-4279} \and \\
Jennifer Burghardt\inst{2}\orcidID{0009-0008-8223-8309} \and 
Eric Rudolph\inst{1}\orcidID{0009-0003-0615-4780} \and 
Jens Albrecht\inst{1}\orcidID{0000-0003-4070-1787}}

\authorrunning{P.~Steigerwald et al.}

\institute{%
Technische Hochschule Nürnberg Georg Simon Ohm\\
Faculty of Computer Science\textsuperscript{1}, 
Institute for E-Counselling\textsuperscript{2}\\
90489 Nuremberg, Germany\\
\email{\{philipp.steigerwald, jennifer.burghardt, \\eric.rudolph, jens.albrecht\}@th-nuernberg.de}\\
}

\maketitle

\begin{abstract}
Text-based online counselling scales across geographical and stigma barriers, yet faces practitioner shortages, lacks non-verbal cues and suffers inconsistent quality assurance.
Whilst artificial intelligence offers promising solutions, its use in mental health counselling raises distinct ethical challenges.
This paper analyses three AI implementation approaches\---autonomous counsellor bots, AI training simulators and counsellor-facing augmentation tools.
Drawing on professional codes, regulatory frameworks and scholarly literature, we identify four ethical principles\---privacy, fairness, autonomy and accountability\---and demonstrate their distinct manifestations across implementation approaches.
Textual constraints may enable AI integration whilst requiring attention to implementation-specific hazards.
This conceptual paper sensitises developers, researchers and practitioners to navigate AI-enhanced counselling ethics whilst preserving human values central to mental health support.

\end{abstract}

\section{Introduction} 

Mental health disorders affect an estimated 970 million people globally, a burden that has grown since the COVID-19 pandemic \cite{kestel_world_2022,torous_evolving_2025,woodward_scalability_2023}.
Text-based online counselling (TBOC)\---delivered synchronously via live-chat or asynchronously via secure email\---has emerged as a promising first-line intervention, enabling vulnerable populations to access support despite stigma-, cost- and geography-related barriers \cite{borghouts_barriers_2021,lattie_overview_2022,yaphe_text-based_2011}.

Yet TBOC faces operational constraints.
Demand is rising faster than professional capacity, leading to increased reliance on volunteers and peer supporters who, whilst valuable, cannot substitute for qualified counsellors \cite{de_boer_forecasting_2023,agyeman-manu_prioritising_2023}.
This workforce gap combines with inherent limitations\---TBOC lacks non-verbal cues for emotional assessment and enables easier disengagement, particularly in asynchronous formats \cite{navarro_exploring_2020}.

These textual constraints, however, make TBOC particularly well-suited for artificial intelligence integration, especially large language models (LLMs) \cite{jin_applications_2025}.
Unlike traditional face-to-face counselling, TBOC already generates structured textual data that AI systems can readily process.
The absence of gestural and vocal complexity means that AI can work with the complete conversational content available to human counsellors.
By leveraging these existing textual records, AI systems may help TBOC realise its full potential\---analysing conversation patterns, identifying risks and supporting both counsellors and counsellees more effectively.
Three established implementation approaches have emerged to address different aspects of TBOC's challenges.
\textbf{Autonomous counsellor bots} engage directly with counsellees, potentially addressing workforce shortages.
\textbf{AI counsellee simulators} provide risk-free training environments for developing clinical competencies.
\textbf{Counsellor-facing augmentation tools} support practitioners through automated documentation and response drafting.
Although AI can be implemented in various other ways, we focus on these three approaches.

Each approach offers potential benefits\---expanded reach, improved consistency, enhanced responsiveness\---yet processes sensitive personal data during moments of vulnerability.
This combination requires ethical analysis beyond generic AI governance.
Developers, researchers and practitioners must balance innovation with protection, efficiency with authenticity, scale with safety.

Current ethical scholarship has primarily examined autonomous chatbots whilst giving less attention to simulators and augmentation tools.
Simulators and augmentation tools receive far less ethical scrutiny than chatbots, despite each AI role presenting distinct challenges that require tailored governance approaches.
We address this gap through three contributions:

\begin{itemize}
    \item A systematic taxonomy of three AI implementation approaches, examining how each reconfigures the counselling relationship through analysis of representative systems.
    \item Identification of four core ethical principles\---privacy, fairness, autonomy and accountability\---that emerge consistently across regulatory frameworks, professional guidelines and scholarly literature.
    \item Application of these principles to each implementation approach, showing how universal requirements manifest differently based on AI's specific role, revealing pattern-specific challenges that may require tailored governance.
\end{itemize}

AI integration in TBOC presents both opportunities and responsibilities\---the potential to expand mental health support alongside obligations to preserve counselling's human elements.
This conceptual paper provides a structured foundation for navigating these considerations.
Rather than dismissing AI due to ethical challenges or adopting it uncritically, we suggest informed dialogue that considers both innovation and vulnerability\---enabling responsible AI deployment in mental health support.

\section{Related Work}\label{sec:related_work}

This section traces the evolution of ethical discourse from foundational digital mental health concerns through artificial intelligence applications to the under-examined intersection of AI and TBOC.

\subsection{Digital Mental Health Ethics: Foundational Concerns}

Early ethical frameworks for digital mental health established core principles that remain relevant today.
Initial narrative reviews of online psychotherapy identified three fundamental ethical concerns: confidentiality breaches, practitioner competence verification and crisis management protocols \cite{ragusea_suggestions_2003}.
Subsequent empirical surveys expanded this foundation, revealing additional challenges specific to text-based modalities\---particularly the absence of non-verbal cues and complications arising from cross-jurisdictional practice \cite{richards_online_2013}.
Comprehensive syntheses of digital intervention ethics confirmed that privacy, competence and safety constitute the ethical foundation across diverse professional guidelines \cite{torous_mental_2018}.
As digital platforms evolved beyond simple communication tools, ethical frameworks necessarily expanded to encompass data governance structures and algorithmic bias mitigation\---recognising that digital systems introduce novel risks beyond traditional therapeutic concerns \cite{pulat_online_2021}.

\subsection{AI Integration in Mental Health: Emerging Ethical Challenges}

The introduction of artificial intelligence into mental health services marked a paradigm shift in ethical complexity.
Machine learning diagnostics and personalised recommender systems transformed passive digital tools into active decision-making agents, intensifying existing debates around privacy whilst introducing new concerns about fairness and explainability \cite{garriga_machine_2022,olawade_enhancing_2024}.
Contemporary large-scale trials exemplify this tension: whilst demonstrating significant clinical benefits, these studies simultaneously reveal fundamental risks stemming from biased training data and opaque algorithmic decision-making \cite{heinz_randomized_2025,thakkar_artificial_2024}.
Consequently, the research community now demands robust safeguards\---including transparent model documentation, representative training datasets and continuous performance monitoring\---as prerequisites for trustworthy AI deployment in mental health contexts \cite{li_security_2023,norori_addressing_2021}.
Regulatory bodies have begun codifying these requirements.
The EU AI Act explicitly mandates comprehensive data governance and bias mitigation for high-risk applications, whilst WHO guidance emphasises continuous human oversight and clear accountability structures in AI-enabled health systems \cite{european_parliament_and_council_of_the_european_union_regulation_2024,world_health_organization_ethics_2021}.

\subsection{Ethical Challenges in AI-Enhanced Text-Based Online Counselling}

TBOC presents unique ethical challenges that AI integration substantially amplifies.
The absence of non-verbal communication fundamentally alters counselling dynamics, complicating rapport building, emotional assessment and\---critically\---crisis identification \cite{yaphe_text-based_2011}.
When AI systems mediate these already-constrained interactions, additional ethical complexities emerge.
Users frequently develop fundamental misconceptions about system capabilities, attributing anthropomorphic understanding to pattern-matching algorithms.
Communicating inherent AI uncertainty becomes particularly challenging in text-only formats.
Traditional counselling boundaries blur when algorithms influence therapeutic conversations \cite{khawaja_your_2023,coghlan_chat_2023}.
Empirical studies reveal persistent implementation gaps: algorithmic bias remains inadequately addressed, data privacy protections lack specificity for therapeutic contexts and informed consent procedures fail to account for AI's invisible influence on counselling interactions \cite{blas_taxonomy_2024}.

Despite these multifaceted challenges, ethical scholarship has disproportionately focused on autonomous chatbot systems, leaving other AI implementation patterns\---particularly training simulators and counsellor-facing augmentation tools\---largely unexamined \cite{ni_scoping_2025}.
This narrow focus proves problematic because different AI roles within counselling workflows generate fundamentally distinct ethical risk profiles requiring tailored governance approaches.
The present study directly addresses this critical gap by systematically examining ethical considerations across all three major AI implementation approaches in TBOC, providing the comprehensive analysis necessary for responsible deployment.

\section{AI in Text-Based Counselling}  
Artificial intelligence presents significant opportunities to scale TBOC delivery and address ongoing capacity limitations. When AI systems enter the traditional counsellor–counsellee dyad, they alter role expectations, conversational dynamics and ethical considerations \cite{jobin_global_2019,meadi_exploring_2025}.

\begin{figure}[htbp]
\centering
\begin{subfigure}[t]{0.3\linewidth}
\centering
\includegraphics[width=\linewidth]{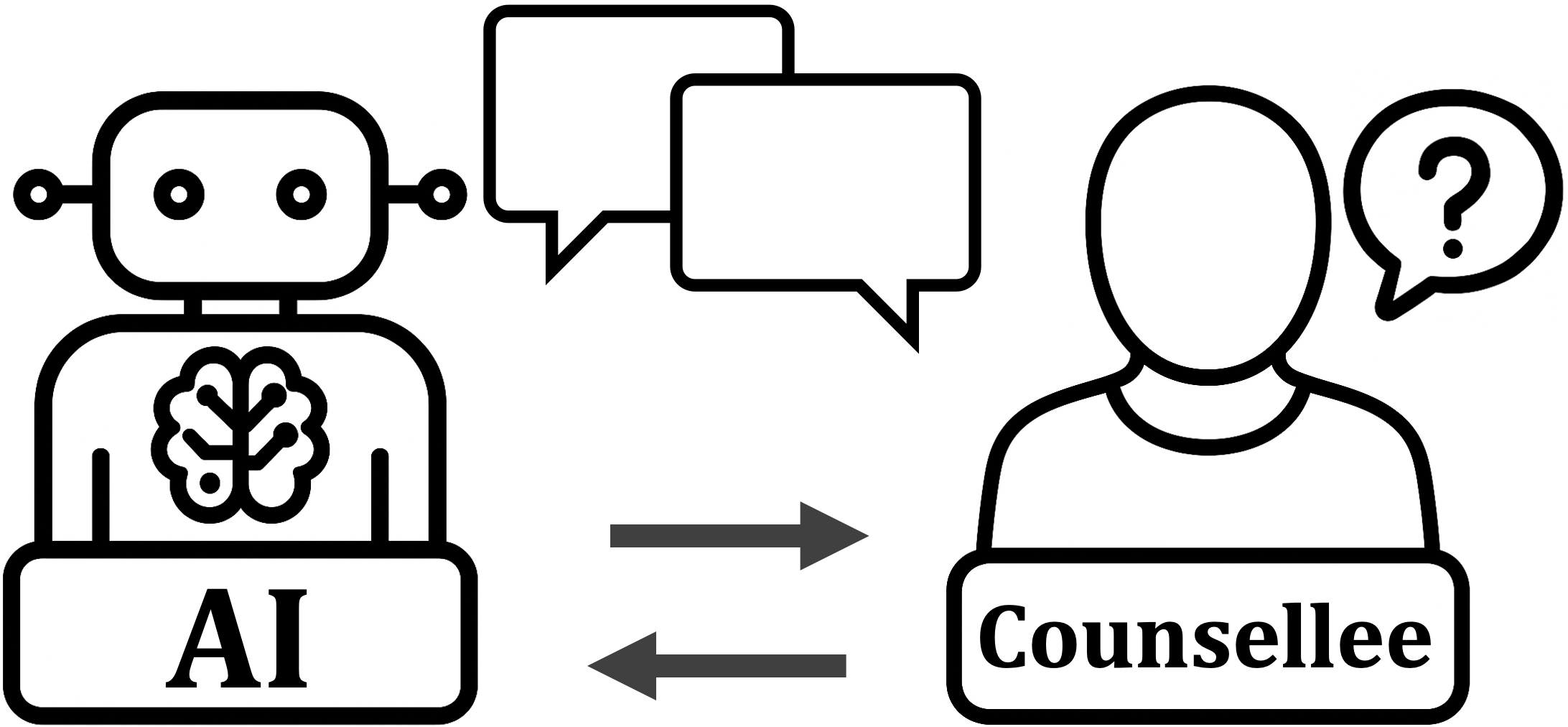}
\caption{Autonomous counsellor or companion bot}
\label{fig:ai_autonomous_counsellor}
\end{subfigure}\hfill
\begin{subfigure}[t]{0.3\linewidth}
\centering
\includegraphics[width=\linewidth]{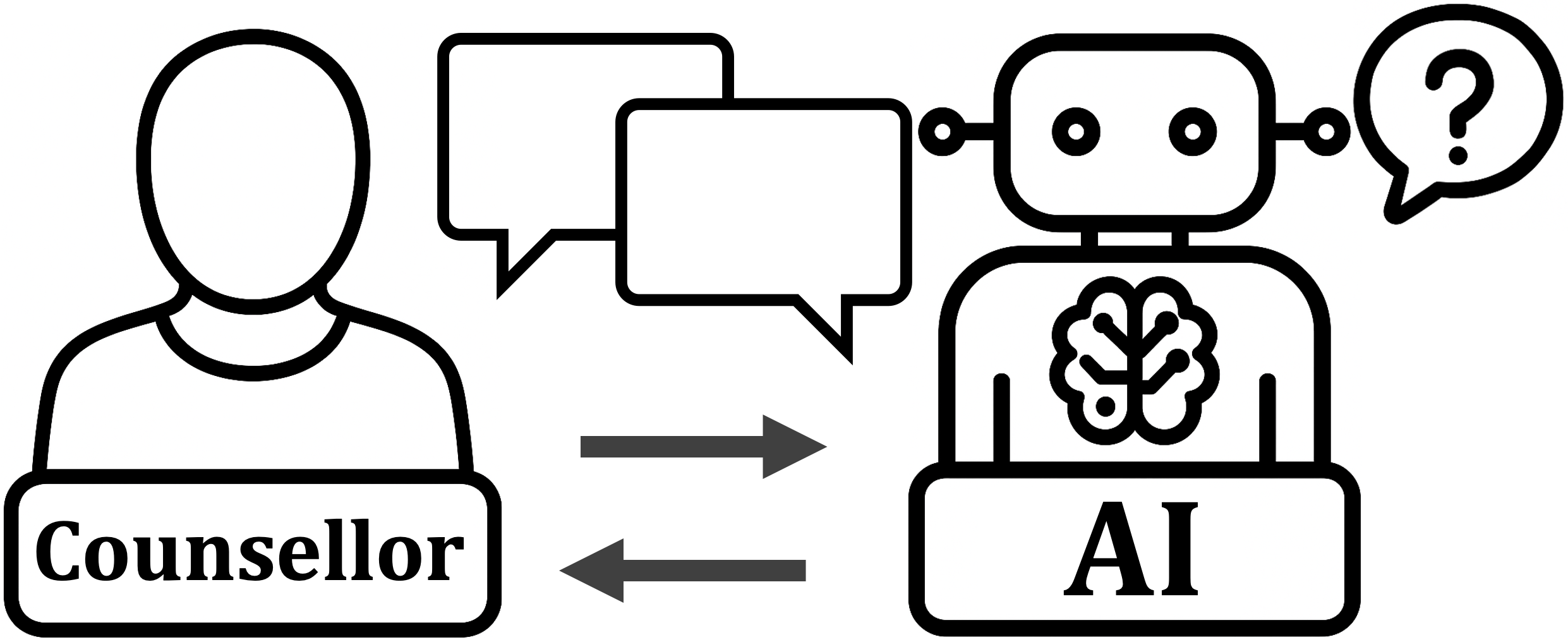}
\caption{AI counsellee simulator}
\label{fig:ai_counsellee_simulator}
\end{subfigure}\hfill
\begin{subfigure}[t]{0.3\linewidth}
\centering
\includegraphics[width=\linewidth]{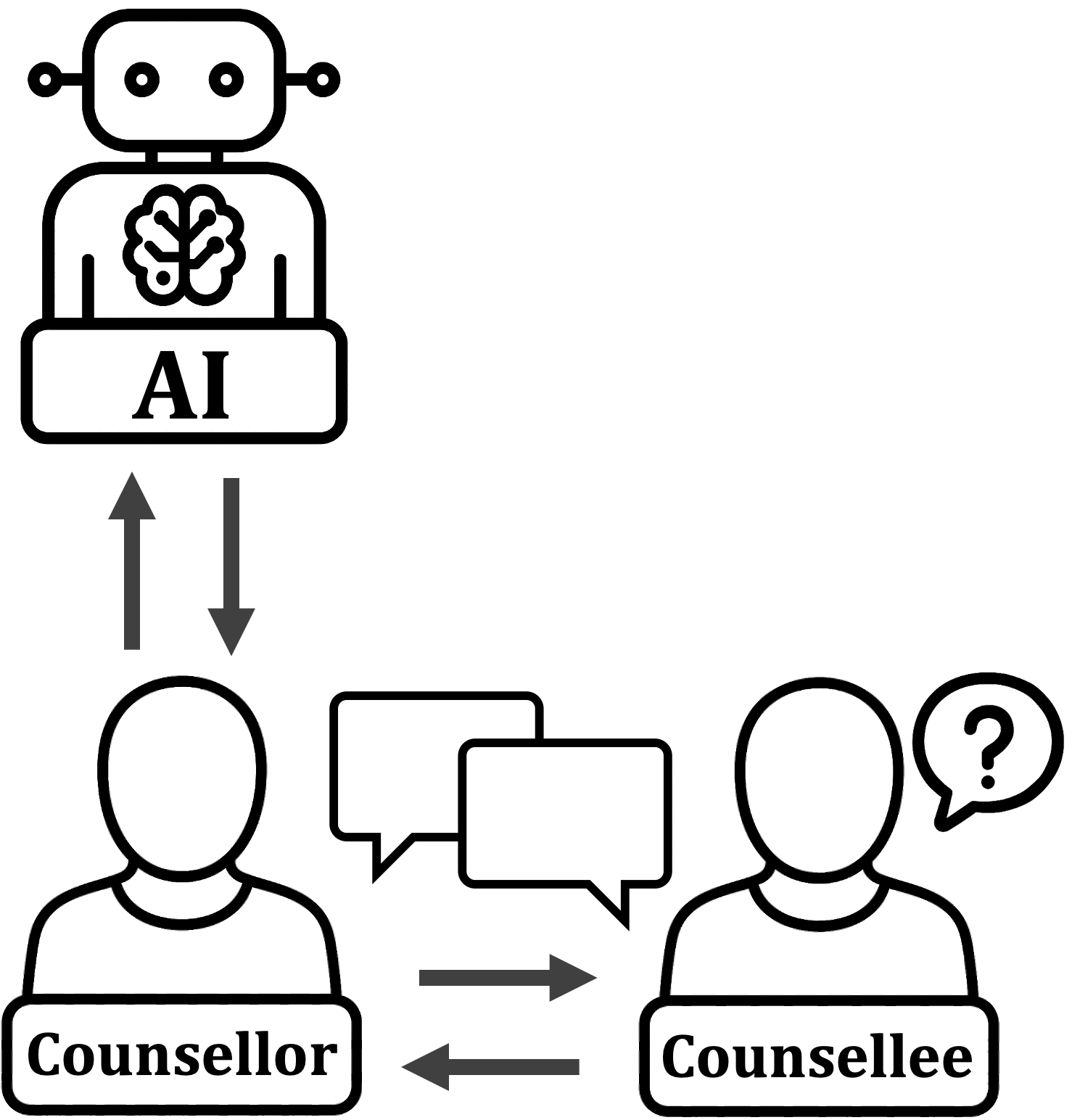}
\caption{Counsellor-facing augmentation}
\label{fig:ai_counsellor_augmentation}
\end{subfigure}
\caption{Placement of three AI implementation approaches within the counselling dyad. 
Arrows indicate principal information flows.}
\label{fig:three_approaches}
\end{figure}

While the literature proposes many configurations for embedding AI into the counsellor–counsellee dyad, we concentrate on three approaches that recur persistently in empirical studies and development reports (Figure~\ref{fig:three_approaches}).

\begin{itemize}
    \item \textbf{Autonomous counsellor or companion bots} function as stand-alone counselling agents that engage directly with counsellees in place of human counsellors. 
    These systems conduct counselling conversations independently, taking primary responsibility for dialogue management, safety protocols and counselling effectiveness.
    \item \textbf{AI counsellee simulators} serve as artificial counsellees that replicate authentic counsellee behaviours for professional training purposes. 
    These systems provide trainees with realistic practice opportunities to develop interviewing and counselling skills in a risk-free environment without the ethical concerns of working with actual counsellees.
    \item \textbf{Counsellor-facing augmentation tools} serve as AI assistants that enhance counsellor performance without altering the counsellor–counsellee communication.
    These systems operate exclusively within the counsellor's workspace, providing real-time support through risk alerts, draft responses and automated documentation.
\end{itemize}

Each implementation approach creates a distinct distribution of responsibilities between human and artificial agents in the counselling process.
This systematic taxonomy of three implementation approaches represents our first contribution, providing a structured approach to understanding how AI reconfigures traditional counselling relationships in fundamentally different ways.
The remainder of this section describes each approach in turn, illustrating typical functionality with representative systems and empirical evidence.

\subsection{AI as Autonomous Counsellor and Companion Systems} 
Chatbots that act as counsellors or well-being companions substitute the human counsellor with a software agent, fundamentally altering the traditional counsellor-counsellee dyad. By replacing the human professional, these systems transform the counselling relationship from a human-to-human interaction into a human-to-machine dialogue.
The lineage reaches back to ELIZA, a rule-based Rogerian script that mirrored user statements through simple pattern matching \cite{weizenbaum_elizacomputer_1966}.
Modern systems deliver psycho-educational content and guided exercises, carrying direct responsibility for conversational safety.

Full-replacement counsellor systems conduct complete interventions without real-time clinical oversight.
Woebot showed significant reductions in depression and anxiety in a two-week trial with college students, though investigators held equity stakes in the vendor \cite{fitzpatrick_delivering_2017}.
Wysa demonstrated effectiveness in a 2024 study where company employees served as co-authors \cite{gao_ai_2024}.
Other Cognitive Behavioral Therapy (CBT) oriented agents include Youper, which yielded symptom gains in an acceptability study \cite{mehta_acceptability_2021} and Fido, which improved depressive outcomes in a Polish randomized controlled trial \cite{karkosz_effectiveness_2024}.
ChatCounsellor represents an open-source LLM fine-tuned on 8,000 counselling transcripts \cite{liu_chatcounselor_2023}.
Tess provides psycho-education via SMS and web chat while terminating sessions when users express suicidal intent \cite{fulmer_using_2018}.
A 2024 survey associated Replika use with lower loneliness and suicidal ideation among 1,006 university users, though qualitative work reveals that safety guardrails can abruptly curtail sensitive dialogue, frustrating users \cite{maples_loneliness_2024,siddals_it_2024}.

Companion-and-referral systems act as first points of contact rather than stand-alone counsellors.
These systems screen distress, offer brief self-help and forward cases to qualified humans when competence limits are reached.
This approach lowers entry barriers while preserving professional primacy.
Limbic Access automates National Health Service self-referrals and forwards structured risk reports to clinicians \cite{rollwage_using_2023}.
Cross-lingual systems like Dr CareSam operate in English and Korean using GPT-4, routing complex queries to human counsellors \cite{kang_development_2025}.
Population-specific variants target particular groups.
ChatPal promotes well-being among rural Europeans in multiple languages, ending daily check-ins with local mental health service menus \cite{potts_multilingual_2023}.
Friend provides psychological first aid to war-affected Ukrainians via Telegram \cite{spytska_use_2025}.

Both full-replacement and companion systems remain highly visible to counsellees, proactive in dialogue and momentarily accountable for safeguarding, making autonomous bots the highest-stakes implementation within our taxonomy.

\subsection{AI Counsellee Simulators for Professional Training} 
AI counsellee simulators enable trainees to practice counselling skills by engaging with AI-controlled simulations of counsellees that maintain fixed personas and specific problem contexts. 
These systems conduct realistic counselling conversations with trainees, addressing several key training challenges in TBOC. 
Unlike traditional role-play approaches that depend on scheduling coordination with human partners, simulators provide unlimited availability for practice-oriented and self-directed training. 
The technology enables trainers to create diverse case examples representing various demographics, mental health conditions and crisis scenarios while maintaining consistent personas across multiple sessions. 
Most importantly, trainees gain independence for self-paced skill development without putting real counsellees at risk \cite{albrecht_virtual_2024}.

The concept traces back to PARRY, a 1972 rule-based program that emulated a paranoid patient for psychiatric interviewing drills \cite{colby_turing-like_1972}.
Contemporary simulators employ LLMs and affective conditioning to produce varied, lifelike dialogue while preserving the turn-taking rhythm of genuine text counselling \cite{albrecht_virtual_2024,wang_patient-_2024}.

Current systems demonstrate substantial pedagogical value.
VirCo maintains stable digital personas and achieved 83\% clinical plausibility in supervisor ratings.
Trainees refined their reflections and summaries over three practice consultations \cite{rudolph_ai-based_2024}.
Patient-\textPsi employs diverse patient cognitive models based on Cognitive Behavioral Therapy principles, allowing trainees to practice formulating cognitive models of patients through role-playing therapy sessions.
This approach yielded an 18\% improvement in case-conceptualisation accuracy in a controlled laboratory study \cite{wang_patient-_2024}.

These simulator systems create genuinely risk-free rehearsal spaces while achieving linguistic and emotional realism.
They expose trainees to diverse personas, supporting deliberate practice, immediate analytics and safe skill acquisition.
Counsellee simulator platforms therefore provide counsellor education with a scalable, data-rich complement to traditional live supervision while preserving counsellee safety and pedagogical rigour.

\subsection{AI Augmentation Systems for Counsellors} 
Augmentation systems operate exclusively in the counsellor’s workspace: counsellees converse with a human, while LLM back-ends unobtrusively generate editable response drafts, risk alerts or analytic feedback that streamline professional work \cite{ni_scoping_2025,blease_generative_2024}.  

An example is CAIA, which extracts and condenses information from lengthy email threads and produces analytical insights, methodological guidance and structured case summaries to support professional counsellors \cite{steigerwald_enhancing_2024}.
During live conversations HAILEY injects micro-suggestions that increase conversational empathy by 19.6\% in a randomised controlled trial with 300 peer supporters on the TalkLife platform \cite{sharma_humanai_2023}.  
CARE delivers real-time motivational-interviewing suggestions to peer counsellors during practice chats, acting as a live co-pilot \cite{hsu_helping_2023}.  
Training-oriented utilities such as Reply+, which rewrites draft messages to match a chosen interpersonal stance, help practitioners refine style without exposing counsellees to experimental phrasing \cite{fu_enhancing_2023}.  

Augmentation tools support counsellors without replacing their judgment\---all clinical decisions remain under human control while the AI operates invisibly to counsellees. This configuration preserves professional autonomy yet creates new ethical obligations for transparency, accountability and bias surveillance.


\subsection{Emerging Hybrid and Transitional Approaches} 
\label{sec:emerging_patterns}

Beyond the three established implementation approaches, new configurations are emerging that challenge traditional assumptions about the counsellor\---counsellee dyad in diverse ways.

Beyond augmentation tools for counsellors, counsellee-support systems represent a parallel development that mirrors augmentation but operates exclusively for the counsellee's benefit.
On-the-fly translation and auto-summarisation run at the counsellee's request whilst remaining invisible to counsellors \cite{institut_fur_e-beratung_ki_2024,kim_mindfuldiary_2024}. 
This asymmetry challenges traditional assumptions about transparency in counselling relationships. 
A different configuration emerges with integrated companion platforms, which transcend simple referral by accompanying counsellees through multiple care phases. 
These systems actively participate in transitions to professional counselling and remain available during formal sessions, blurring boundaries between formal and informal support. 
Moving toward more radical reconfigurations, triadic counselling represents a true hybrid approach, introducing AI as a visible third actor that creates an "artificial third" presence \cite{haber_artificial_2024}. 
These manifest as proactive co-facilitators shaping dialogue or on-demand assistants activated when needed. 
Such configurations fundamentally alter power dynamics and transference patterns in ways current theory cannot address. 
Finally, post-session reflection systems take yet another approach by analysing completed sessions to generate feedback for counsellors or insights for counsellees \cite{rudolph_automated_2024}. 
Operating outside the counselling interaction preserves authenticity whilst enabling data-driven improvement.

These diverse approaches\---asymmetric support tools, integrated companions, triadic hybrids and post-session analytics\---demonstrate that autonomy, visibility and role allocation represent deliberate design choices rather than fixed constraints.
Each configuration generates distinct ethical challenges.
Asymmetric systems create information imbalances, continuous companions blur consent boundaries, triadic relationships disrupt confidentiality frameworks and post-session analytics complicate data ownership and temporal consent.
Although beyond our current scope, these emerging patterns demand urgent attention.
They represent not incremental improvements but fundamental reimaginings of how AI might participate in counselling relationships.
As these models mature, they will require entirely new ethical frameworks\---ones that navigate the complex interplay between human vulnerability and algorithmic capability.

\section{Ethical Challenges of AI in Text-Based Online Counselling} 

AI integration in mental health counselling introduces fundamental ethical complexities stemming from both the sensitive nature of counselling conversations and inherent counsellee vulnerability.
This section first establishes four core ethical principles derived from regulatory and professional guidance, then systematically demonstrates how these principles manifest distinctly across autonomous counsellor bots, AI counsellee simulators and counsellor-facing augmentation tools.

\subsection{General Ethical Challenges in AI-Enhanced Text-Based Online Counselling} 
\label{sec:general_ethics}

Despite apparent fragmentation across disciplines, comprehensive analysis of professional codes, regulatory frameworks and scholarly literature reveals convergence around four core ethical principles: \textbf{privacy}, \textbf{fairness}, \textbf{autonomy} and \textbf{accountability}.

The APA Ethical Principles and Code of Conduct protects privacy (Standard 4.01), fairness (Standard 3.01, Principle D), informed consent (Standard 3.10, Principle E) and accountability (Principle B, Standards 2.01, 3.04) \cite{american_psychological_association_ethical_2017}. 
WHO's 2021 guidance covers privacy and autonomy (Principles 1–2), accountability through safe deployment (Principle 4) and bias mitigation (Principle 5) \cite{world_health_organization_ethics_2021}, with its 2024 addendum providing operational checkpoints for privacy (§ 4.2), autonomy and informed consent (§ 4.3), bias mitigation (§ 4.4) and accountability (§ 4.7) \cite{world_health_organization_ethics_2025}. 
Key literature streams\---research on AI codes of practice \cite{jobin_global_2019,floridi_unified_2019}, chatbot scoping reviews \cite{saeidnia_ethical_2024,meadi_exploring_2025} and counselling position papers \cite{kieslinger_wenn_2024,vilaza_is_2021}\---consistently identify these same four principles. 
The EU Artificial Intelligence Act reinforces this framework through privacy and data governance (Articles 10–11), fairness via bias-controlled training data (Article 10(3)), autonomy through human-in-control clauses (Articles 13–14) and accountability (Articles 17, 61–65) \cite{european_parliament_and_council_of_the_european_union_regulation_2024}.

Each principle addresses specific harms in TBOC.  
Privacy breaches expose sensitive mental health data, algorithmic bias creates discriminatory barriers, opacity undermines counsellee agency and unclear accountability leaves vulnerable users without recourse.
Together, these principles establish the ethical foundation for responsible AI deployment in TBOC.

\paragraph{Privacy} requires robust data governance extending beyond basic encryption. 
Whilst end-to-end encryption secures transmission, AI systems must decrypt and process actual content to generate responses, creating vulnerability points where sensitive counselling data exists unencrypted within AI processing environments. 
Organisations must ensure secure channels to AI services, implement safeguards against data leaks during processing and establish encrypted return paths for AI-generated responses.
Additionally, metadata patterns can reveal access behaviours and crisis frequencies even with encryption. 
Training data requires active consent and rigorous anonymisation that eliminates identifying traces whilst preserving counselling patterns, with automatic data purging when retention periods expire.

\paragraph{Fairness} requires continuous bias mitigation throughout system lifecycles.
All AI inherits training data biases\---the challenge lies in distinguishing legitimate patterns from harmful discrimination.
Intersectional gaps prove particularly problematic: groups marginalised across ethnicity, gender, socioeconomic status and linguistic variation face compounded risks of misdiagnosis or exclusion.
Static testing proves inadequate, deployed systems demand continuous monitoring as populations and language evolve.

\paragraph{Autonomy} necessitates transparent AI disclosure and meaningful user control.
Counsellees must understand when AI shapes their counselling experience and retain authority to modify or override generated content.
Consent must extend beyond binary acceptance to granular, revocable permissions governing data collection, processing and sharing.
Critically, counsellees must access human-only interaction without penalties or restrictions.

\paragraph{Accountability} establishes traceable responsibility spanning technical systems and human oversight.
Organisations must designate clear decision authority for each component, maintaining escalation pathways when AI proves insufficient.
Developers retain duty-of-care obligations beyond traditional liability, whilst explainability becomes essential for trust and attribution.
Comprehensive frameworks require external audits, incident protocols and transparent reporting mechanisms.

Additional considerations\---transparency, beneficence, non-maleficence, justice, trust, explicability and environmental sustainability\---appear throughout the literature \cite{jobin_global_2019,floridi_unified_2019,laine_ethics-based_2024,radanliev_ai_2025,coghlan_chat_2023}.
These typically elaborate the four core principles or vary across contexts.
Subsequent sections address these where relevant whilst maintaining focus on the four principles demonstrating consistent prioritisation across regulatory and professional guidance.

\subsection{Autonomous Counsellor and Companion Bots} 
\label{sec:autonomous_bots}

Amongst the three implementation approaches, autonomous systems bear the heaviest ethical burden.
Every utterance and decision originates solely from the machine, without human mediation.
Whilst these systems promise to address practitioner shortages and reduce stigma barriers, their direct counsellee interaction generates ethical complexities that fundamentally exceed those of other AI implementation approaches.

\paragraph{Privacy} vulnerabilities multiply through direct counsellee interaction.
Unlike invisible augmentation tools, autonomous bots must explicitly request consent whilst maintaining counselling rapport\---a delicate balance that human counsellors navigate intuitively but machines must algorithmically approximate.
Crisis communications demand immediate data segregation with enhanced encryption, yet comprehensive protection conflicts with the system's need to detect escalating crises, forcing explicit trade-offs between confidentiality and safety.

\paragraph{Fairness} becomes life-critical when machines make unsupervised triage decisions.
Performance disparities across demographic groups translate directly into inappropriate crisis escalations.
Cultural variations in expressing distress\---via metaphor, indirection or silence\---challenge systems trained predominantly on Western datasets.
Unlike other AI applications where bias causes inconvenience, fairness failures in autonomous counselling can prove fatal.

\paragraph{Autonomy} requires systems to identify as artificial whilst maintaining counselling engagement.
Autonomous systems must recognise their limitations and actively facilitate transitions to human care when counsellee needs exceed algorithmic capabilities.
This demands sophisticated boundary detection that seamlessly connects counsellees to appropriate services without creating abandonment experiences.
Counsellees must access human professionals immediately upon request, with systems facilitating rather than obstructing these transitions.

\paragraph{Accountability} becomes exceptionally complex when machines make independent counselling decisions.
Traditional liability frameworks assume human judgment; autonomous systems disrupt these assumptions entirely.
Clear protocols must specify transfer triggers with redundant fail-safes ensuring handoffs during system failures.
Liability structures must address foreseeable failure modes: unrecognised crises, culturally inappropriate responses, false reassurance in dangerous circumstances.

Beyond these principles, autonomous systems confront insurmountable limitations.
Professional competence gaps cannot be bridged\---no training data substitutes for clinical education and supervised practice.
Crisis response capabilities remain critically constrained: systems cannot summon emergency services, conduct binding safety plans or initiate involuntary interventions.
The authenticity paradox deepens when sophisticated empathy simulation fosters genuine emotional dependence on systems incapable of reciprocal care.
Boundary management requires acknowledging limitations whilst avoiding both counselling overstatement and unnecessary referral delays.

These compounding challenges establish autonomous systems as the highest-risk AI implementation approaches in TBOC.
Deployment demands not merely technical sophistication but fundamental reconceptualisation of counselling ethics and regulatory frameworks.
The path forward requires honest recognition that autonomous systems cannot replace human counsellors but might\---with sufficient safeguards\---provide valuable first-line support whilst facilitating access to human care.

\subsection{AI Counsellee Simulators for Training} 
\label{sec:simulators}

When artificial agents replace real counsellees in training contexts, the ethical landscape fundamentally transforms.
The primary concern shifts from protecting vulnerable counsellees to ensuring educational integrity and professional preparedness\---a reorientation that generates distinct ethical challenges.

\paragraph{Privacy} requires data governance operating across dual axes\---safeguarding trainee learning records whilst ensuring synthetic personas cannot be traced to real individuals.
Training data must undergo thorough anonymisation and pseudonymisation to prevent reproduction of actual counselling sessions or counsellee identification.
Even aggregated patterns require careful transformation to prevent re-identification whilst maintaining clinical authenticity.
Additionally, trainee performance data\---including practice transcripts and skill assessments\---must remain confidential, accessible to educators only through explicit agreements and protected from unauthorised access by employers or licensing bodies.
The tension persists\---creating sufficiently authentic narratives for effective training whilst eliminating all privacy-compromising traces.

\paragraph{Fairness} manifests through representational accuracy and assessment equity.
Synthetic personas must reflect population demographics without over- or underrepresenting groups whilst avoiding reductive stereotypes.
The challenge lies in distinguishing harmful stereotypes from legitimate clinical patterns\---systems must represent genuine counselling dynamics without reinforcing discriminatory assumptions.
Training scenarios require sufficient variance to accommodate diverse counselling approaches across cultural contexts.
Assessment algorithms must evaluate multiple counselling styles rather than imposing singular paradigms.
Educational bias proves particularly insidious when homogeneous training shapes entire counsellor cohorts, potentially marginalising equally valid alternative methodologies.

\paragraph{Autonomy} requires ensuring trainees understand they interact with AI systems in an educational context, not with real counsellees.
Trainees must understand they engage with synthetic entities designed for skill development, not genuine counselling relationships.
This transparency enables pedagogically valuable behaviours\---experimentation, deliberate mistake-making, unlimited scenario repetition\---that would violate ethical boundaries in actual counselling.
Trainees require meaningful control over difficulty levels, competency areas and progression readiness.

\paragraph{Accountability} centres on educational validity rather than counselling effectiveness.
The crucial question becomes whether synthetic interactions transfer meaningfully to real counselling relationships.
Current validation studies often measure simulator performance rather than subsequent real-world competence.
Governance must specify how personas are created, validated, updated and retired to maintain educational relevance whilst preventing harmful representations from persisting.

Beyond these reconceptualised principles, simulators confront unique challenges.
The fundamental problem of ecological validity questions whether synthetic interactions adequately prepare trainees for genuine counselling's unpredictability and emotional intensity.
Trainees may master scripted scenarios whilst remaining unprepared for actual counsellee complexity and ambiguity.
Educational dependency emerges when predictable response patterns impair ability to navigate real counsellee contradictions.
Temporal considerations create extended vulnerabilities\---training data persisting years beyond completion could influence career trajectories through credentialing reviews or liability proceedings.

These accumulating challenges demand careful positioning of simulators in counsellor education.
Rather than replacing human-to-human training, simulators might best serve as preparatory tools building foundational competencies before actual counsellee engagement.
This acknowledges both risk-free practice value and the irreplaceable nature of genuine human interaction in developing counselling expertise.
Only through honest recognition of pedagogical limitations alongside training benefits can simulators contribute meaningfully without compromising professional preparedness or perpetuating systemic biases.

\subsection{Counsellor-Facing Augmentation Tools}\label{sec:augmentation_tools}

When AI systems operate invisibly within the practitioner's workspace\---summarising transcripts, flagging risks, suggesting responses\---the ethical focus shifts to mediated consent, professional discretion and institutional oversight.
This fundamental invisibility to counsellees creates ethical challenges distinct from direct AI-counsellee interactions.

\paragraph{Privacy} confronts the complexity of invisible processing.
Unlike autonomous bots where counsellees knowingly engage with AI, augmentation tools analyse conversations without explicit counsellee awareness.
Counsellees may reasonably assume their communications remain within traditional human-only relationships, creating heightened privacy obligations.
The challenge lies in maintaining AI functionality whilst ensuring counsellees understand algorithmic processing occurs behind the scenes.
Transient processing with minimal retention becomes essential when explicit AI consent is absent.

\paragraph{Fairness} centres on invisible influence over counselling quality.
Systematic AI biases may subtly shape counsellor responses across demographic groups without either party recognising this influence.
Unlike observable bias in direct AI interactions, augmentation bias operates through human intermediaries, complicating detection and mitigation.
Monitoring must track both AI performance and shifts in counsellor behaviour patterns across counsellee populations.
Counsellors risk unknowingly perpetuating AI biases whilst believing they exercise independent professional judgment.

\paragraph{Autonomy} must balance professional discretion with bilateral informed consent.
Counsellors need genuine control over AI suggestions without succumbing to automation bias\---where time pressure or over-reliance produces uncritical acceptance.
Counsellees deserve informed choice about AI involvement, yet excessive disclosure about invisible processes could disrupt counselling relationships.
Creating meaningful consent that respects both professional autonomy and counsellee choice without compromising efficacy remains challenging.

\paragraph{Accountability} must navigate shared human-AI responsibility.
Augmentation tools create hybrid accountability structures that differ from autonomous systems' primary responsibility and simulators' educational accountability.
Counsellors retain ultimate clinical responsibility whilst AI invisibly influences professional judgment through potentially untraceable pathways.
Clear protocols must specify documentation, review and override procedures for AI suggestions to maintain professional accountability.

Beyond these principles, invisible operation creates distinct challenges.
Professional skill atrophy emerges when counsellors unconsciously become AI-de\-pen\-dent, gradually eroding clinical judgment.
Counsellors may gradually develop excessive reliance on AI support without recognising this dependency.
Time pressure during sessions exacerbates automation bias, encouraging uncritical acceptance of AI suggestions.
This hidden influence undermines authenticity\---counsellees assume purely human care whilst AI invisibly shapes their counselling experience.

Augmentation demands careful attention to consent, professional boundaries and clinical judgment preservation.
Success requires balancing AI transparency with relationship integrity.
These tools cannot replace human expertise but might\---with appropriate safeguards\---enhance counselling without compromising its human essence.

\section{Conclusion}

Text-based online counselling serves as a crucial first-line intervention, enabling vulnerable populations to access mental health support despite barriers of stigma, cost and geography.
This paper explored how artificial intelligence might support TBOC whilst maintaining its ethical foundation.
The analysis revealed three key insights.
First, examining autonomous counsellor bots, AI training simulators and augmentation tools showed that each implementation approach reshapes the counselling relationship in distinct ways.
Second, whilst privacy, fairness, autonomy and accountability remain consistent ethical concerns, their practical implications vary depending on how AI is deployed.
Third, each implementation brings unique challenges\---autonomous systems struggle with crisis response and competence boundaries, simulators risk perpetuating stereotypes and may not transfer to real practice, whilst augmentation tools raise concerns about professional dependency and the ethics of invisible AI mediation.
By sparking critical discourse, this work raises awareness that whilst LLMs can transform TBOC, what we build and how we implement it will determine whether technology amplifies or undermines counselling's human essence.

Several limitations constrain this analysis.
The literature reviewed primarily draws from European and North American contexts, which may not fully capture diverse global perspectives on AI ethics in counselling.
Additionally, this conceptual analysis would benefit from empirical validation through stakeholder consultations and longitudinal studies.
Future research should incorporate cross-cultural perspectives and develop standardised evaluation metrics for evidence-based deployment.

The constraints that limit TBOC may also create opportunities for AI integration.
Large language models can process extensive conversation histories, maintain consistency across sessions and provide round-the-clock availability\---areas where human counsellors face practical limitations.
The textual nature of TBOC data becomes an enabling condition for AI deployment.
With appropriate safeguards, AI could help address these constraints.
Autonomous systems might provide immediate support when humans are unavailable, simulators could expand training access across geographical barriers and augmentation tools may free counsellors to focus on relational depth.

As a conceptual paper establishing foundational ethical frameworks, we advocate for thoughtful engagement with AI and LLMs in TBOC.
Beyond these frameworks lies a simple truth\---we must expand mental health support to meet unprecedented demand.
The question isn't whether to use AI or LLMs in counselling, but how to deploy them responsibly whilst maintaining rigorous ethical standards.
TBOC provides that opportunity\---available today, scalable tomorrow, with the potential to unite human expertise and ethical AI deployment for comprehensive mental health support at scale.
In text-based counselling, we have found an unexpected gateway where human expertise and artificial intelligence converge to help people more quickly and efficiently.
This is not about replacing human connection\---it's about creating more pathways to it.

\bibliographystyle{splncs04}
\bibliography{clean_references_FAIEMA}
\end{document}